# Nonreciprocal Pancharatnam-Berry Metasurface for Unidirectional Wavefront Manipulation

Hao Pan[1], Mu Ku Chen[2], Din Ping Tsai[2], and Shubo Wang[1,3] *

[1]Department of Physics, City University of Hong Kong, Tat Chee Avenue, Kowloon, Hong Kong, China.
[2]Department of Electrical Engineering, City University of Hong Kong, Tat Chee Avenue, Kowloon, Hong Kong, China.
[3]City University of Hong Kong Shenzhen Research Institute, Shenzhen, Guangdong 518057, China.

*Corresponding author: shubwang@cityu.edu.hk

**ABSTRACT**

Optical metasurfaces have been widely used for manipulating electromagnetic waves due to their low intrinsic loss and easy fabrication. The metasurfaces employing the Pancharatnam-Berry (PB) geometric phase, called PB metasurfaces, have been extensively applied to realize spin-dependent functionalities, such as beam steering, focusing, holography, etc. The demand for PB metasurfaces in complex environments has brought about one challenging problem, i.e., the interference of multiple wave channels that limits the performance of PB metasurfaces. A promising solution is developing nonreciprocal PB metasurfaces that can isolate undesired wave channels and exhibit unidirectional functionalities. Here, we propose a mechanism to realize nonreciprocal PB metasurfaces of subwavelength thickness by using the magneto-optical effect of YIG material in synergy with the PB geometric phase of spatially rotating meta-atoms. Using full-wave numerical simulations, we show that the metasurface composed of dielectric cylinders and a thin YIG layer can achieve nearly 92% and 81% isolation of circularly polarized lights at 5.5 GHz and 6.5 GHz, respectively, attributed to the enhancement of the magneto-optical effect by the resonant Mie modes and Fabry-Pérot cavity mode. In addition, the metasurface can enable efficient unidirectional wavefront manipulations of circularly polarized lights, including nonreciprocal beam steering and nonreciprocal beam focusing. The proposed metasurface can find highly useful applications in optical communications, optical sensing, and quantum information processing.

# I. INTRODUCTION

The recent decade has witnessed significant progress in the design and fabrication of artificial optical structures working at microwave [1,2], terahertz [3,4], infrared [5,6], and visible optical bands [7,8], which can exhibit intriguing electromagnetic (EM) properties not existing in nature [9,10]. One important type is an ultrathin layer of structures known as metasurfaces. Metasurfaces can induce strong light-matter interaction in the nanoscale, and they benefit from small intrinsic loss and easy fabrication compared to conventional bulky metamaterials. By carefully designing the subwavelength elements (i.e., meta-atom) in each unit cell, metasurfaces can give rise to various fascinating wavefront-manipulation functionalities, such as perfect absorption [11-13], structural colors [14,15], anomalous reflection or refraction [16-18], surface wave excitation [19,20], metalens [21-23], metaholograms [24-26], and many others [27,28]. Different from conventional diffractive optical devices, metasurfaces not only can employ the resonance phase and propagation phase (also called dynamic phase) but also can utilize the PB geometric phase derived from the spatial rotation of meta-atoms [29-33], leading to the so-called PB metasurfaces. Therefore, the PB metasurfaces can acquire an extra degree of freedom to control the wavefront of circularly polarized (CP) light besides the resonance and dynamic phases, giving rise to many intriguing phenomena such as photonic spin Hall effect [34,35], vortex beam generation [36,37], etc. Meanwhile, the PB metasurfaces can serve as a powerful platform for developing CP-light-associated applications, e.g., the CP wave control of the motions of biomolecules exhibiting chiral structures [38,39]. Consequently, the PB metasurfaces have great application potential in the next-generation photonic devices with multifunctionalities.

Despite the PB metasurfaces' unprecedented performance in wavefront manipulation, their functionalities are intrinsically restricted by the Lorentz reciprocity [40]. Introducing additional mechanisms to break the reciprocity of PB metasurfaces can generate new functionalities that are essential to many applications, such as invisible sensing, full-duplex communication, and noise-tolerant quantum computation, etc., where nonreciprocity can prevent the backscattering from defects or boundaries [41-43]. In addition, nonreciprocity allows metasurfaces to exhibit different properties for the opposite propagating waves, thus giving rise to Janus-type functionalities. One effective way to achieve nonreciprocity is by using gyroelectric or gyromagnetic materials sensitive to magnetic-field biasing, which exhibit asymmetric permittivity or permeability tensor accounting for the Faraday-magneto-optical (FMO) effect [44]. Qin *et al.* proposed a set of self-biased nonreciprocal magnetic metasurfaces to achieve bidirectional wavefront modulation based on the different hybrid resonant-dynamic phase profiles for bidirectional CP waves [45]. However, the local resonant phase can be easily affected by the coupling from neighboring meta-atoms, resulting in an undesired phase profile that limits the performance of the metasurface. In contrast, the PB geometric phase only determined by the rotation angle of meta-atoms is a better mechanism for realizing the stable nonreciprocal wavefront manipulation. Zhao *et al.* presented an interesting metadevice combining a PB metasurface and an anisotropic metasurface, which can simultaneously realize phase modulation and nonreciprocal isolation [46]. The metadevice involves a complex multilayer structure with a large thickness that inevitably affects its diffraction efficiency. Therefore, simple and thin PB metasurfaces capable of achieving high-efficiency nonreciprocal wavefront manipulation are highly desirable.

In this article, we report a nonreciprocal PB metasurface composed of elliptical dielectric cylinders and a thin YIG layer to simultaneously realize PB-phase-based wavefront manipulation and microwave isolation. The thin YIG layer under a magnetic field bias can give rise to strong spin-selective isolation due to the FMO effect and Fabry-Pérot (FP) resonance. Meanwhile, the resonant coupling between the Mie modes in the dielectric cylinder and the FP mode in the YIG layer can effectively tune the nonreciprocal band by enhancing the FMO effect, thus achieving a high isolation ratio of nearly 92% and 81% at 5.5 GHz and 6.5 GHz, respectively. The PB phase of each meta-atom can be individually controlled via the corresponding optical-axis rotation and is unaffected by the FMO effect. Following a digital coding metasurface design methodology, the proposed nonreciprocal PB metasurfaces can offer multiple functionalities with high isolation, which are demonstrated by the tailored metadeflector with nonreciprocal beam steering and the metalens with nonreciprocal focusing. Our proposed all-dielectric nonreciprocal PB metasurface can find applications in multiple fields, e.g., EM wave isolation, nonreciprocal antennas, optical sensing, quantum information processing, etc.

## II. RESULTS AND DISCUSSIONS
### A. Unidirectional spin-selective nonreciprocal metasurface

The metasurface consists of subwavelength meta-atoms arranged in a square lattice with period $l$, as depicted in Fig. 1(a). Each meta-atom comprises a dielectric cylinder sitting on a YIG substrate of thickness $t$. Under the external biased magnetic field along the +$z$-direction, the YIG is characterized by a permeability tensor with asymmetric off-diagonal elements [44]

$$\mu = \mu_0 \begin{bmatrix} \mu_r & i\kappa_r & 0 \\ -i\kappa_r & \mu_r & 0 \\ 0 & 0 & 1 \end{bmatrix},  \tag{1}$$

where $\mu_r = 1 + \dfrac{\omega_0 \omega_m}{\omega_0^2 - \omega_m^2}$, $\kappa_r = \dfrac{\omega \omega_m}{\omega_0^2 - \omega^2}$, $\omega_m = \mu_0 \gamma M_s$, $\omega_0 = \mu_0 \gamma H_0 + i\omega\alpha$, $\gamma = 1.579 \times 10^{11}$ C/kg is the gyromagnetic ratio, $4\pi M_s = 1780$ G is the saturation magnetization, $B_0 = \mu_0 H_0 = 0.05$ T is the external magnetic field, $\alpha = 0.002$ is the damping factor, and $\mu_0$ is the vacuum permeability. The relative permittivity of YIG is $\varepsilon_{r1} = 15$. The dielectric cylinder has relative permittivity $\varepsilon_{r2} = 24$ and relative permeability $\mu_{r2} = 1$. Its major and minor axis are $a$ and $b$, respectively, and its height is $h$. The orientation of the cylinder is denoted by the angle $\theta$. In this configuration, the metasurface exhibits different refractive indices for the incident CP waves with the wavevector parallel and antiparallel to the biased magnetic field, attributed to the FMO effect. Thus, the metasurface can give rise to spin- and direction-dependent manipulation of EM waves.

To illustrate the physical mechanism, we first analyze the spin-selective transmission of normally incident CP waves in the thin infinite YIG layer with the $+z$ magnetic field bias. The reflection and transmission coefficients for the left-hand circularly polarized (LCP) and right-hand circularly polarized (RCP) EM waves with wavevectors antiparallel to the biased magnetic field can be derived straightforwardly (See Appendix A for the derivations), and their expressions are

$$r_{RL}^{+f} = \frac{(1 - Y_L^{f\,2})(e^{i2k_L^f t} - 1)}{(Y_L^f + 1)^2 e^{i2k_L^f t} - (Y_L^f - 1)^2}, \quad t_{LL}^{+f} = \frac{4 Y_L^f e^{i(k_L^f + k_0)t}}{(Y_L^f + 1)^2 e^{i2k_L^f t} - (Y_L^f - 1)^2},$$

$$r_{LR}^{+f} = \frac{(1 - Y_R^{f\,2})(e^{i2k_R^f t} - 1)}{(Y_R^f + 1)^2 e^{i2k_R^f t} - (Y_R^f - 1)^2}, \quad t_{RR}^{+f} = \frac{4 Y_R^f e^{i(k_R^f + k_0)t}}{(Y_R^f + 1)^2 e^{i2k_R^f t} - (Y_R^f - 1)^2}, \tag{2}$$

where $Y_L^f = \sqrt{\varepsilon_{r1}/\mu_r + \kappa_r}$ and $Y_R^f = \sqrt{\varepsilon_{r1}/\mu_r - \kappa_r}$ are the relative wave admittances for the LCP and RCP waves forward propagating in the YIG layer, $k_L^f = \sqrt{\varepsilon_{r1}(\mu_r + \kappa_r)} k_0$ and $k_R^f = \sqrt{\varepsilon_{r1}(\mu_r - \kappa_r)} k_0$ are the corresponding LCP and RCP wavevectors in the YIG layer, and $k_0$ is the wavevector in free space. Here, the superscript "+" denotes the $+z$-biased magnetic field, "$f$" denotes forward incidence (i.e., $-z$-direction), and the subscript "$RL$" ("$LR$") denotes the CP conversion from LCP to RCP (RCP to LCP). It can be noted from Eq. (2) that the off-diagonal element $\kappa_r$ in the permeability tensor results in the different impedances and wavevectors for the LCP and RCP waves, which leads to the differences in the co-polarized transmission ($|t_{LL}^{+f}|^2$ and $|t_{RR}^{+f}|^2$) and cross-polarized reflection ($|r_{RL}^{+f}|^2$ and $|r_{LR}^{+f}|^2$). Figure 1(b) shows the transmission spectra given by Eq. (2) (denoted by the blue symbol lines). For the considered thin YIG layer, the FP cavity resonance can enhance the FMO effect and increase the transmission difference (i.e., $|t_{RR}^{+f}|^2 - |t_{LL}^{+f}|^2$) for the LCP and RCP incident waves. This transmission difference can reach a maximum of nearly 94% around 6.8 GHz. To verify the analytical results, we conducted full-wave finite-element simulations by using COMSOL and computed the transmission spectra. The numerical results are denoted by the red symbol lines in Fig. 1(b), which is consistent with the analytical results. In addition, as the forward (−z-direction) normally incident LCP wave is equivalent to the backward (+z-direction) normally incident RCP wave for the infinite YIG layer with the +z-biased magnetic field, Fig. 1(b) also indicates that the thin YIG layer can exhibit an evident nonreciprocal-transmission feature, i.e., the large transmission contrast between the forward and backward CP waves (See Appendix A for details).

Figure 1(c) shows the transmission spectra of the PB metasurface under the incidence of the LCP and RCP plane waves. We set $\theta=0°$ for the rotation angle of all the dielectric cylinders. Due to the breaking of cylindrical symmetry by the meta-atoms, the helicity of the wave is not conserved, and the transmitted wave generally contains both LCP and RCP components. We notice that the transmission is dominated by the cross-polarized components $t_{RL}^{+f}$ and $t_{LR}^{+b}$ for the forward LCP and backward RCP incidence, respectively, which have two resonance peaks at 5.5 GHz and 6.5 GHz with differences ($|t_{RL}^{+f}|^2 - |t_{LR}^{+b}|^2$) of 92% and 81%, respectively. The large isolation ratio can be attributed to the FMO effect of the YIG layer enhanced by the Mie resonance in the dielectric cylinders. To understand the effect of the Mie resonant modes of the cylinders, we show in Fig. 1(d) the numerically calculated multipole decomposition of the cylinder scattering power under the excitation of the forward incident LCP wave (See Appendix C for the multipole decomposition). It is noted that the two resonances at 5.5 GHz and 6.5 GHz are mainly attributed to the magnetic dipole mode and the hybrid magnetic dipole-electric quadrupole mode, respectively. The resonant electric and magnetic field amplitudes are shown in the insets of Fig.1(d). We notice that the magnetic field inside the cylinder is strongly enhanced at 5.5 GHz, while both the electric and magnetic fields are strongly localized in the cylinder at 6.5 GHz due to the resonant electric quadrupole and magnetic dipole resonances. The resonant coupling between these hybrid Mie resonances and the FP cavity resonance in the YIG layer can enhance the interaction between the wave and the magnetic material, leading to the enhanced FMO effect and thus the strong nonreciprocity of the metasurface [47].

We further investigate the relationship between the nonreciprocal properties of the PB metasurface and various system parameters, including the cylinder height $h$, the biased magnetic field $B_0$, and the incident angle. Figure 2(a) shows the numerically simulated isolation ratio $|t_{RL}^{+f}|^2 - |t_{LR}^{+b}|^2$ as a function of the cylinder height $h$ for the system in Fig. 1(a). As seen, the isolation peaks undergo redshift as $h$ increases, which is expected since the eigenfrequencies of the Mie modes in the cylinder are generally inversely proportional to the geometric size of the cylinder. Specifically, as $h$ varies, the spectral profile of the first resonance maintains a Lorentz shape, where the local maximum of the isolation remains above 85%. In contrast, the spectral profile of the second resonance undergoes dramatic variation due to the interference with other multipoles, as evidenced by the sharp transition of isolation from negative to positive values.

Figure 2(b) shows the isolation ratio $|t_{RL}^{+f}|^2 - |t_{LR}^{+b}|^2$ of the proposed metaisolator when the external magnetic field is $B_0 = 0.05$ T, $B_0 = 0$ T, and $B_0 = -0.05$ T (corresponding to red, magenta, and blue symbol lines, respectively). We notice that the results for different biasing directions are nearly antisymmetric with respect to the case of $B_0 = 0$ T which induces zero isolation. This can be understood as follows. The transmission coefficients follow the relationships $t_{RL}^{+f} = t_{RL}^{+b}$ and $t_{LR}^{+b} = t_{LR}^{+f}$ because the forward normally incident LCP (RCP) wave is converted to RCP (LCP) wave by the elliptical cylinder and the resulting RCP (LCP) wave is equivalent to the LCP (RCP) wave backward normally incident on the YIG layer (similar to the property of single YIG layer mentioned above). In addition, the magnetic field bias direction decides the spin-selective transmission of the metaisolator. For the opposite magnetic biasing, we can obtain the relationships $t_{RL}^{+b} = t_{LR}^{-b}$ and $t_{LR}^{+f} = t_{RL}^{-f}$ (See Appendix B for details). Consequently, we have the relationships $t_{RL}^{+f} = t_{LR}^{-b}$ and $t_{LR}^{+b} = t_{RL}^{-f}$, and thus $|t_{RL}^{+f}|^2 - |t_{LR}^{+b}|^2 = -(|t_{RL}^{-f}|^2 - |t_{LR}^{-b}|^2)$, i.e., reversing the direction of biased magnetic field leads to a sign change of the isolation value in Fig. 2(b). Figure 2(c) shows the dependence of the isolation on the magnitude of the external magnetic field. We notice that the isolation peaks at 5.5 GHz and 6.5 GHz are blue-shifted without obvious reduction of the isolation ratio, demonstrating the robust performance of the proposed metasurface isolator.

We also investigate the effect of the incident angle of CP waves on the isolation. At large incident angles, higher-order diffractions can appear, and we only consider the isolation for the $0^{th}$-order cross-polarized transmission under the forward LCP and backward RCP wave incidence with the same incident angle. As depicted in Fig. 2(d), the isolation at the resonance frequency of 5.5 GHz will slightly shift with the increase of the incident angle. At large incident angles, the isolation at 5.5 GHz is reduced owing to the combined effect of the resonance shift and change of CP conversion efficiency in the elliptical cylinder. Notably, the isolation can still reach above 80% for the incident angle as large as 45°. Interestingly, the isolation ratio at 6.5 GHz is insensitive to the variation of the incident angle, and it can maintain a large value above 80% for the incident angle within [0°, 60°]. Therefore, the proposed nonreciprocal metaisolator can achieve a stable and high isolation ratio at the targeted frequencies for a wide range of incident angles, which lays the foundation for further nonreciprocal wavefront manipulations.

In addition to manipulating the wave amplitude, the metasurface can also be applied to achieve unidirectional phase manipulation for the transmitted CP wave. This is done by varying the orientational angle $\theta$ of the dielectric cylinder to induce PB geometric phases, as shown by the inset in Fig. 3. For CP waves normally forward incident on the metasurface, the output waves can be expressed as

$$\begin{bmatrix} E_L^{out} \\ E_R^{out} \end{bmatrix} = \begin{bmatrix} 0 & t_{LR}^{\pm f} e^{-i2\theta(x,y)} \\ t_{RL}^{\pm f} e^{i2\theta(x,y)} & 0 \end{bmatrix} \begin{bmatrix} E_L^{in} \\ E_R^{in} \end{bmatrix}, \quad (3)$$

where $t_{LR}^{\pm f}$ and $t_{RL}^{\pm f}$ are the cross-polarized transmission coefficients for the forward incident RCP and LCP waves, respectively. The superscript "±" denotes the direction of the external biased magnetic field $B_0$. The dielectric cylinder can induce a PB phase shift $\varphi = 2\sigma\theta$, where $\sigma = +1$ ($\sigma = -1$) for the LCP (RCP) wave. Figure 3 shows the simulated amplitude of the transmitted electric field (blue symbol line) and the PB phase (red symbol line) for different orientation angles of the cylinder. As seen, the orientation angle $\theta$ of the cylinder has a negligible impact on the transmission amplitude, which is around 96% for different rotation angles. Meanwhile, the PB phase agrees with the relationship $\varphi = 2\sigma\theta$. The stable high CP transmission and the PB phase of $2\pi$ range lay the foundation for designing wavefront-manipulation metasurfaces.

### B. Nonreciprocal PB metadeflector for beam steering

Owing to the superior nonreciprocal isolation under the large-angle incidence and the stable PB phase of the meta-atoms, it is possible to construct a nonreciprocal metadeflector with an on-demand phase profile to manipulate the propagation direction of the incident CP beam. Figure 4(a) schematically shows the concept of the nonreciprocal PB-phase-based metadeflector with the +z-biased magnetic field. The meta-atoms are invariant along $y$ direction, but they are orientated differently in the $x$ direction to induce the PB geometric phase profile.

At 5.5 GHz, the metasurface can convert the forward incident LCP wave into the RCP wave and deflect it away from the normal direction. Meanwhile, the metasurface can isolate the backward RCP wave incident along the opposite deflection direction, i.e., the time-reversed wave of the deflected RCP wave.

The transmitted wavevector and the incident wavevector satisfy the phase-matching condition in the periodic structure [48]:

$$k_{out} = k_{in} + m k_{PB}, \qquad (4)$$

where $k_{out} = 2\pi \sin\theta_{out}/\lambda$, $k_{in} = 2\pi \sin\theta_{in}/\lambda$, $k_{PB} = 2\pi/P$, $\theta_{in}$ and $\theta_{out}$ are the incident and deflected angles, respectively, $\lambda$ is the incident wavelength, $P$ is the period size of the supercell (covering $2\pi$ phase range) along the $y$-direction, and $m$ is the deflection order. For the normally incident wave ($\theta_{in} = 0°$), Eq. (4) can be simplified as $\sin\theta_{out} = m\lambda/P$ where the supercell period $P = Nl$ with $N$ being the meta-atom number in the supercell and $l$ being the meta-atom period. The discrete PB phase profile in the supercell can be expressed as $\varphi(n) = 2\pi n/N$ where $n$ denotes the $n$-th meta-atom in the supercell, thus requiring a rotation angle distribution $\theta(n) = \pi n/N$. Following this principle, we design four different metadeflectors working at 5.5 GHz with the supercells consisting of 4, 6, 8, and 12 meta-atoms, respectively. These metasurfaces induce the 1st-order diffraction at the angles 74.64°, 40°, 28.82°, and 18.75°, respectively. Figure 4(b) shows the simulated electric field ($E_y$) profiles at 5.5 GHz for the four metasurfaces. The deflection angles of the output beam are consistent with analytical values given by Eq. (4). Under the forward normal incidence, the 1st-order diffraction efficiency in these four cases is 66.44%, 92.1%, 95.59%, and 96.08%, respectively. Under the backward incidence, the transmission efficiency in the four cases is 6.64%, 0.79%, 0.026%, and 0.002%, respectively. Accordingly, the isolation ratios are 59.8%, 91.31%, 95.564%, and 96.078%, which demonstrate the highly efficient nonreciprocal beam steering function of the proposed metadeflectors. Additionally, we note that for 6-, 8-, and 12-cell cases, the deflected beams are mainly composed of the 1st-order diffraction, while higher-order diffraction components begin to appear in the output beam of the 4-cell case, which can be attributed to the large wavevector component parallel to the metasurface. The emergence of the higher-order diffractions in this case decreases the isolation ratio and leads to a complex output wavefront.

### C. Nonreciprocal PB metalens for beam focusing

The PB-phase-based planar metalenses with excellent performance, e.g., high numerical aperture (NA), have been widely proposed and fabricated, generating broad applications in imaging [49,50], microscopy [51], and spectroscopy [52,53]. However, the effect of backscattering is usually neglected in conventional PB metalenses, thus limiting their applications in the platforms requiring anti-echo and anti-reflection functions. Introducing nonreciprocity to PB metalenses can be a solution to this problem. This corresponds to the concept of nonreciprocal PB metalens for unidirectional beam focusing, as illustrated in Fig. 5(a). The forward normally incident LCP wave passes through the nonreciprocal metalens with the +z-biased magnetic field and is focused into one spot, but the RCP wave radiated from the focusing spot, i.e., the time-reversal excitation, will be blocked by the metalens, thus realizing the nonreciprocal beam focusing.

The PB phase profile $\varphi(x,y)$ of the metalens should follow [49]

$$\varphi(x,y) = \frac{2\pi}{\lambda}\left(f - \sqrt{x^2 + y^2 + f^2}\right), \tag{5}$$

where $\lambda$ is the wavelength, $f$ is the focal length, $x$ and $y$ are the coordinates of each meta-atom. Similar to the metadeflector mentioned above, we consider the metalens with invariant phase profile in $x$-direction. The rotation angle profile of the meta-atoms in this case is $\theta(y) = \frac{\pi}{\lambda}\left(f - \sqrt{y^2 + f^2}\right)$, which has the discretized form $\theta(n) = \frac{\pi}{\lambda}\left(f - \sqrt{n^2 l^2 + f^2}\right)$, where $n$ denotes the $n$-th meta-atom, and $l$ is the period of each meta-atom. To demonstrate the nonreciprocal focusing functionality, we design three metalenses with different focal lengths 1.5$\lambda$, 2$\lambda$, and 3$\lambda$ ($\lambda$=54.5 mm at 5.5 GHz), respectively. We conduct numerical simulations for the nonreciprocal focusing realized by the three metalenses. Figure 5(b) depicts the simulated electric-field distributions in the $yz$-plane with the forward incident LCP (the upper panels) and the backward RCP radiation from the focal point (the bottom panels). It is noticed that the forward incident LCP waves are focused into spots at different focal points. The corresponding focal lengths are determined to be 80.62 mm, 108.85 mm, and 149.83 mm, respectively. The discrepancy between the theoretical and simulated focal lengths can be attributed to the coupling effect between the adjacent meta-atoms. Figures 5(c)-(e) show the intensity on the focal planes with the diffraction-limited ($\lambda$/(2×NA)) full width at half-maximum (FWHM) of 30 mm, 30.27 mm, and 31.47 mm, respectively. The corresponding NA of the metalenses is 0.908, 0.9, and 0.866, respectively. To understand the nonreciprocity of the metalenses, we calculate the light transmission under the forward incidence, which reaches 79.9%, 85.1%, and 89.64% for the three cases, respectively. Meanwhile, the focusing efficiency is found to be 68.18%, 73.14%, and 74.51% for the three cases, respectively, where the focusing efficiency is defined as the fraction of the incident light that passes through a circular aperture in the focal plane with a diameter equal to three times of the FWHM spot size [54]. Additionally, we find that the backward RCP radiation from the focal point only gives rise to the transmission of 12.9%, 14.67%, and 10.9%, respectively. Therefore, the isolation ratios of the three metalenses are 55.28%, 58.47%, and 63.61%, respectively. The contrast between the focusing efficiency under forward incidence and the transmission under backward radiation clearly demonstrates the nonreciprocal focusing functionality of the designed PB metalenses.

### III. CONCLUSION

To summarize, we have demonstrated that high-performance nonreciprocal wavefront manipulation of CP beams can be achieved by using the magnetic-biased PB metasurfaces consisting of elliptical dielectric cylinders and a thin magnetic YIG layer. Due to the strong resonant coupling between the Mie modes in the cylinders and the FP cavity mode in the thin YIG layer, the FMO effect can be greatly enhanced near the resonant frequencies, thus giving rise to significant spin-selective nonreciprocal isolation. Meanwhile, the stable PB phase and the large isolation ratio over a wide range of incident angles can guarantee efficient nonreciprocal wavefront manipulation. By designing the PB phase gradient profile, we have demonstrated two types of nonreciprocal functional metasurfaces: the metadeflectors that can realize nonreciprocal beam steering with different deflection angles, and the high-NA metalenses that can realize nonreciprocal focusing with different focal lengths. The proposed nonreciprocal PB metasurfaces can simultaneously achieve high-efficiency wavefront manipulation and large isolation ratio, which pave the way to the applications in wave multiplexing for high-capacity communications and optical imaging with anti-reflection functions.

### ACKNOWLEDGEMENTS

The work described in this paper was supported by grants from the Research Grants Council of the Hong Kong Special Administrative Region, China (Projects No. AoE/P-502/20 and No. CityU 11308223).

### APPENDIX A: SPIN-SELECTIVE TRANSMISSION OF AN INFINITE YIG LAYER

The yttrium iron garnet (YIG) material is a common magnetic material that can show obvious asymmetry spin characteristics, e.g., the different propagation constants and impedances for the orthogonal CP states, due to the large off-diagonal elements in the permeability tensor under the external magnetic field biasing. For the considered thin YIG layer, the FP cavity resonance can enhance the FMO effect. To understand this property, we analytically determine the transmission and reflection for different CP waves propagating through the YIG layer.

Consider an LCP wave backward (+x direction) normally incident onto the YIG layer of thickness $t$ and with the +x-biased magnetic field, as shown in Fig. 6, the electric and magnetic fields in regions 1, 2, and 3 can be expressed as:

Region 1:

$$\mathbf{E}_{inc} = E_0 \begin{pmatrix} 0 \\ 1 \\ i \end{pmatrix} e^{-ik_0 x}, \quad \mathbf{H}_{inc} = Y_0 E_0 \begin{pmatrix} 0 \\ -i \\ 1 \end{pmatrix} e^{-ik_0 x}, \quad \mathbf{E}_{ref} = \begin{pmatrix} 0 \\ E'_y \\ E'_z \end{pmatrix} e^{ik_0 x}, \quad \mathbf{H}_{ref} = Y_0 \begin{pmatrix} 0 \\ E'_z \\ -E'_y \end{pmatrix} e^{ik_0 x}. \quad (A1)$$

Region 2:

$$\mathbf{E}_{\underset{LCP}{\rightarrow}} = E_{L1} \begin{pmatrix} 0 \\ 1 \\ i \end{pmatrix} e^{-ik_L^b x}, \quad \mathbf{H}_{\underset{LCP}{\rightarrow}} = E_{L1} Y_0 Y_L^b \begin{pmatrix} 0 \\ -i \\ 1 \end{pmatrix} e^{-ik_L^b x}, \quad \mathbf{E}_{\underset{RCP}{\rightarrow}} = E_{R1} \begin{pmatrix} 0 \\ 1 \\ -i \end{pmatrix} e^{-ik_R^b x}, \quad \mathbf{H}_{\underset{RCP}{\rightarrow}} = E_{R1} Y_0 Y_R^b \begin{pmatrix} 0 \\ i \\ 1 \end{pmatrix} e^{-ik_R^b x}, \quad (A2)$$

$$\mathbf{E}_{\underset{LCP}{\leftarrow}} = E_{L2} \begin{pmatrix} 0 \\ 1 \\ -i \end{pmatrix} e^{ik_L^f x}, \quad \mathbf{H}_{\underset{LCP}{\leftarrow}} = E_{L2} Y_0 Y_L^f \begin{pmatrix} 0 \\ -i \\ -1 \end{pmatrix} e^{ik_L^f x}, \quad \mathbf{E}_{\underset{RCP}{\leftarrow}} = E_{R2} \begin{pmatrix} 0 \\ 1 \\ i \end{pmatrix} e^{ik_R^f x}, \quad \mathbf{H}_{\underset{RCP}{\leftarrow}} = E_{R2} Y_0 Y_R^f \begin{pmatrix} 0 \\ i \\ -1 \end{pmatrix} e^{ik_R^f x}. \quad (A3)$$

Region 3:

$$\mathbf{E}_{tran} = \begin{pmatrix} 0 \\ E''_y \\ E''_z \end{pmatrix} e^{-ik_0 x}. \quad (A4)$$

where $Y_0 = \sqrt{\varepsilon_0/\mu_0}$ is the wave admittance in free space, $Y_L^b = Y_R^f = \sqrt{\varepsilon_{r1}/(\mu_r - \kappa_r)}$ and $Y_R^b = Y_L^f = \sqrt{\varepsilon_{r1}/(\mu_r + \kappa_r)}$ are the relative wave admittances for LCP and RCP waves normally forward (-x-direction) and backward (+x-direction) propagating in the YIG material, $k_L^b = k_R^f = \sqrt{\varepsilon_{r1}(\mu_r - \kappa_r)}k_0$ and $k_R^b = k_L^f = \sqrt{\varepsilon_{r1}(\mu_r + \kappa_r)}k_0$ are the wave vectors of the LCP and RCP waves normally forward (-x-direction) and backward (+x-direction) propagating in the YIG material, $k_0$ is the wave vector in free space. Furthermore, according to the boundary conditions ($\mathbf{e_n} \times (\mathbf{H_1} - \mathbf{H_2}) = 0$, $\mathbf{e_n} \times (\mathbf{E_1} - \mathbf{E_2}) = 0$) between regions 1 and 2 ($x=0$), and between regions 2 and 3 ($x=t$), we can get eight equations to solve for the eight unknowns in Eq. (A1-A4), which can be expressed as below:

At $x=0$:

$$\begin{aligned} E_0 + E'_y &= E_{L1} + E_{R1} + E_{L2} + E_{R2} \\ iE_0 + E'_z &= iE_{L1} - iE_{R1} - iE_{L2} + iE_{R2} \\ Y_0(-iE_0 + E'_z) &= -iE_{L1}Y_0 Y_L^b + iE_{R1}Y_0 Y_R^b - iE_{L2}Y_0 Y_L^f + iE_{R2}Y_0 Y_R^f \\ Y_0(E_0 - E'_y) &= E_{L1}Y_0 Y_L^b + E_{R1}Y_0 Y_R^b - E_{L2}Y_0 Y_L^f - E_{R2}Y_0 Y_R^f \end{aligned} \quad (A5)$$

At $x=t$:

$$E_{L1}e^{-ik_L^b t} + E_{R1}e^{-ik_R^b t} + E_{L2}e^{ik_L^f t} + E_{R2}e^{ik_R^f t} = E_y''e^{-ik_0 t}$$

$$iE_{L1}e^{-ik_L^b t} - iE_{R1}e^{-ik_R^b t} - iE_{L2}e^{ik_L^f t} + iE_{R2}e^{ik_R^f t} = E_z''e^{-ik_0 t}$$

$$-iE_{L1}Y_0Y_L^b e^{-ik_L^b t} + iE_{R1}Y_0Y_R^b e^{-ik_R^b t} - iE_{L2}Y_0Y_L^f e^{ik_L^f t} + iE_{R2}Y_0Y_R^f e^{ik_R^f t}$$
$$= -Y_0 E_z'' e^{-ik_0 t}$$

$$E_{L1}Y_0Y_L^b e^{-ik_L^b t} + E_{R1}Y_0Y_R^b e^{-ik_R^b t} - E_{L2}Y_0Y_L^f e^{ik_L^f t} - E_{R2}Y_0Y_R^f e^{ik_R^f t}$$
$$= -Y_0 E_y'' e^{-ik_0 t} \quad (A6)$$

By solving the Eq. (A5-A6), we can get the solutions:

$$E_{L1} = \frac{2E_0(Y_L^b + 1)e^{i2k_L^b t}}{(Y_L^b + 1)^2 e^{i2k_L^b t} - (Y_L^b - 1)^2}$$

$$E_{R1} = 0$$

$$E_{L2} = 0$$

$$E_{R2} = \frac{2E_0(Y_R^f - 1)}{(Y_R^f + 1)^2 e^{i2k_R^f t} - (Y_R^f - 1)^2}$$

$$E_y' = \frac{E_0(1 - Y_L^{b\,2})(e^{i2k_L^b t} - 1)}{(Y_L^b + 1)^2 e^{i2k_L^b t} - (Y_L^b - 1)^2}$$

$$E_z' = i\frac{E_0(1 - Y_L^{b\,2})(e^{i2k_L^b t} - 1)}{(Y_L^b + 1)^2 e^{i2k_L^b t} - (Y_L^b - 1)^2}$$

$$E_y'' = \frac{4E_0 Y_L^b e^{i(k_L^b + k_0)t}}{(Y_L^b + 1)^2 e^{i2k_L^b t} - (Y_L^b - 1)^2}$$

$$E_z'' = i\frac{4E_0 Y_L^b e^{i(k_L^b + k_0)t}}{(Y_L^b + 1)^2 e^{i2k_L^b t} - (Y_L^b - 1)^2} \quad (A7)$$

From Eq. (A7), we can observe that there exists the LCP wave along the +x-direction and the RCP wave along the -x-direction in the YIG layer, thus inducing coherent interference and the FP cavity resonance. Additionally, we note that the reflected and transmissive waves are always the RCP and LCP waves, respectively, and their coefficients can be represented as

$$r_{RL}^{+b} = \frac{(1 - Y_L^{b\,2})(e^{i2k_L^b t} - 1)}{(Y_L^b + 1)^2 e^{i2k_L^b t} - (Y_L^b - 1)^2},$$

$$t_{LL}^{+b} = \frac{4Y_L^b e^{i(k_L^b + k_0)t}}{(Y_L^b + 1)^2 e^{i2k_L^b t} - (Y_L^b - 1)^2}, \quad (A8)$$

where the superscript "+" indicates the +x-directional magnetic biasing, "b" represents the backward normal incidence (+x-direction), "RL" symbolizes the CP state conversion from LCP to RCP, and of course "LL" stands for the CP state conservation for LCP wave. Similarly, we also can get the solutions of the reflected and transmitted coefficients for the case of the RCP wave normally backward (+x-direction) passing through the t-thick YIG layer with +x-directional magnetic biasing:

$$r_{LR}^{+b} = \frac{(1-Y_R^{b\,2})(e^{i2k_R^b t}-1)}{(Y_R^b+1)^2 e^{i2k_R^b t}-(Y_R^b-1)^2},$$

$$t_{RR}^{+b} = \frac{4Y_R^b e^{i(k_R^b+k_0)t}}{(Y_R^b+1)^2 e^{i2k_R^b t}-(Y_R^b-1)^2}. \tag{A9}$$

Comparing Eq. (A8) with Eq. (A9), it can be concluded that the difference in spin-dependent reflection and transmission is determined by the off-diagonal element $\kappa_r$. Furthermore, when the external magnetic field reverses the direction, the off-diagonal element in the permeability tensor will change from $\kappa_r$ to $-\kappa_r$, thus the intrinsic admittance and wave vector of the LCP (RCP) wave in the case of +x-biased magnetic field case will be equal to those of the RCP (LCP) wave in the case of –x-biased magnetic field. Therefore, the corresponding reflection and transmissive coefficients can be expressed by

$$r_{RL}^{+b}=r_{LR}^{-b}, \quad r_{LR}^{+b}=r_{RL}^{-b}$$
$$t_{LL}^{+b}=t_{RR}^{-b}, \quad t_{RR}^{+b}=t_{LL}^{-b} \tag{A10}$$

Meanwhile, due to the symmetry feature of YIG layer relative to the yz-plane, the reverse of the applied magnetic field is equivalent to the reverse of the incident direction of CP waves, thus getting $r_{RL}^{+b}=r_{LR}^{+f}$, $r_{LR}^{+b}=r_{RL}^{+f}$, $t_{LL}^{+b}=t_{RR}^{+f}$, and $t_{RR}^{+b}=t_{LL}^{+f}$. Thus, the spin-selective transmission and reflection also depend on the incident direction in addition to the magnetic field biasing.

### APPENDIX B: DEPENDENCE OF TRANSMISSION ON THE DIRECTIONS OF INCIDENCE AND BIASED MAGNETIC FIELD FOR METAISOLATOR

The relationship between the CP states, wave propagation direction, and the biased magnetic field direction for the proposed metaisolator is numerically verified in Fig. 7. It can be noted in Fig. 7(a) that the cross-polarized transmissions for normally forward and backward incident LCP waves are nearly identical (i.e., $t_{RL}^{+f}=t_{RL}^{+b}$). This is also true for the RCP wave (i.e., $t_{LR}^{+f}=t_{LR}^{+b}$). A similar phenomenon can also be found in the case of −z-biased magnetic field shown in Fig. 7(b). This can be understood as follows. For the forward LCP wave passing through the metasurface with the +z-biased magnetic field, the cross-polarized transmission can be expressed as $t_{RL}^{+f}=t_c^{L\to R}t_R^{+f}$, where $t_c^{L\to R}$ denotes the conversion from LCP wave to RCP wave and $t_R^{+f}$ is the forward RCP transmission for the YIG layer under the resonant coupling of the dielectric cylinder. Similarly, for the backward LCP wave, the cross-polarized transmission can be represented by $t_{RL}^{+b} = t_L^{+b} t_c^{L\to R}$, where $t_L^{+b}$ is the backward LCP transmission for the YIG layer under the resonant coupling of the dielectric cylinder. It should be noted that the forward RCP transmission $t_R^{+f}$ is equal to the backward LCP transmission $t_L^{+b}$ for the YIG layer in the presence of the Mie resonances of the cylinder, similar to property of the single YIG layer (discussed in Appendix A). Since the Mie resonances in the elliptical cylinder are spin-independent, the efficiency of its coupling to the YIG layer is unaffected by the CP state. Therefore, $t_{RL}^{+f} = t_{RL}^{+b}$ can be concluded. Meanwhile, their co-polarized transmission can also be expressed as $t_{LL}^{+f} = (1-t_c^{L\to R})t_L^{+f}$ and $t_{LL}^{+b} = t_L^{+b}(1-t_c^{L\to R})$, respectively. Since $t_L^{+f} \neq t_L^{+b}$ owing to the nonreciprocal characteristic of YIG, we can obtain $t_{LL}^{+f} \neq t_{LL}^{+b}$. Additionally, the magnetic-biased direction determines the spin-selective property of the metaisolator due to the electromagnetic characteristic of YIG. As demonstrated by the equal transmission of different CP states passing through the metaisolators with the opposite magnetic biasing, i.e., $t_{LR}^{+f} = t_{RL}^{-f}$ and $t_{RL}^{+f} = t_{LR}^{-f}$. To summarize, these relationships can be described by $t_{RL}^{+f} = t_{RL}^{+b} = t_{LR}^{-f} = t_{LR}^{-b}$ and $t_{LR}^{+f} = t_{LR}^{+b} = t_{RL}^{-f} = t_{RL}^{-b}$.

## APPENDIX C: ELECTROMAGNETIC MULTIPOLE EXPANSION

The external field can induce the charge density $\rho$ and current density **J** in the metasurface, which give rise to electromagnetic multipoles. Therefore, the resonance response of the metastructure can be understood based on the multipole decompositions. The multipole moments can be evaluated using the current density **J**(**r**) within the unit cell ($\alpha, \beta, \gamma = x, y, z$) as [55-57]:

$$\mathbf{p} = \frac{1}{i\omega} \int \mathbf{J} d^3 r, \tag{C1}$$

$$\mathbf{m} = \frac{1}{2c} \int [\mathbf{r} \times \mathbf{J}] d^3 r, \tag{C2}$$

$$\mathbf{T} = \frac{1}{10c} \int \left[ (\mathbf{r} \cdot \mathbf{J})\mathbf{r} - 2r^2 \mathbf{J} \right] d^3 r, \tag{C3}$$

$$Q_{\alpha,\beta}^e = \frac{1}{2i\omega} \int [r_\alpha J_\beta + r_\beta J_\alpha - \frac{2}{3}\delta_{\alpha,\beta}(\mathbf{r}\cdot\mathbf{J})] d^3 r, \tag{C4}$$

$$Q_{\alpha,\beta}^m = \frac{1}{3c} \left\{ \int [(\mathbf{r}\cdot\mathbf{J})_\alpha r_\beta] d^3 r + \int [(\mathbf{r}\cdot\mathbf{J})_\beta r_\alpha] d^3 r \right\}, \tag{C5}$$

$$Q_{\alpha,\beta}^T = \frac{1}{28c} \int [4r_\alpha J_\beta(\mathbf{r}\cdot\mathbf{J}) - 5r^2(r_\alpha J_\beta + r_\beta J_\alpha) - 2r^2 \delta_{\alpha,\beta}(\mathbf{r}\cdot\mathbf{J})] d^3 r, \tag{C6}$$

where **p**, **m**, **T**, **Q**$^e$, **Q**$^m$, and **Q**$^T$ represent the electric dipole, magnetic dipole, toroidal dipole, electric quadrupole, magnetic quadrupole, and toroidal quadrupole, respectively, $c$ is the light speed. The total scattered power $I_s$ of the metasurface can be expressed as [58]

$$I_s = \frac{2\omega^4}{3c^3}|\mathbf{p}|^2 + \frac{2\omega^4}{3c^3}|\mathbf{m}|^2 + \frac{4\omega^5}{3c^4}(\mathbf{p}\cdot\mathbf{T}) + \frac{2\omega^6}{3c^5}|\mathbf{T}|^2 + \frac{\omega^6}{5c^5}\sum\left|Q^e_{\alpha,\beta}\right|^2 + \frac{\omega^6}{40c^5}\sum\left|Q^m_{\alpha,\beta}\right|^2 + O\left(\frac{1}{c^5}\right) \quad (C7)$$

We evaluated each term on the right-hand side of Eq. (C7) for the metaisolator, and the results are shown in Fig. 1(d).

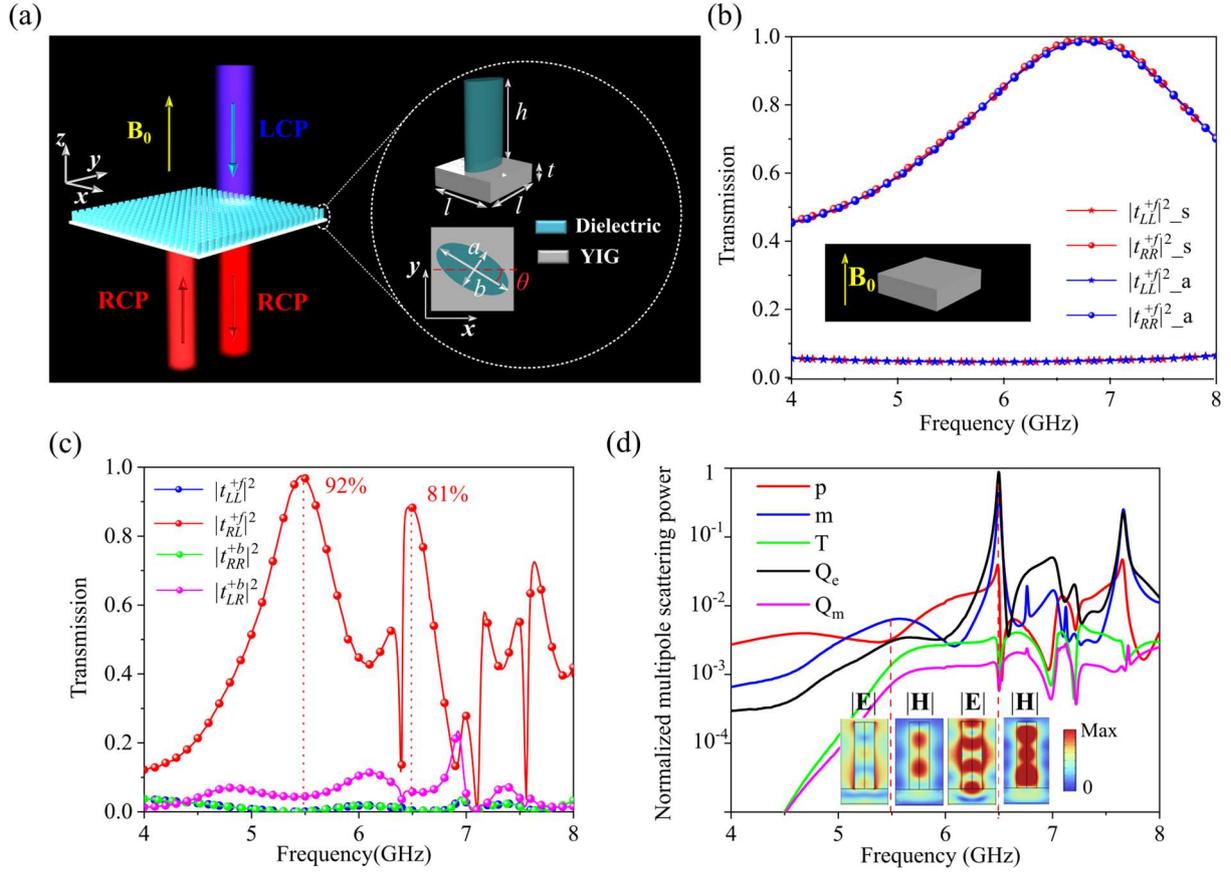

FIG. 1. The PB metasurface isolator and its nonreciprocal properties. (a) The schematics of the metaisolator and the meta-atom. The meta-atom is composed of an elliptical dielectric resonator and a YIG layer with geometric parameters $l$=14 mm, $h$=20 mm, $a$=12 mm, $b$=6 mm, and $t$=4.5 mm. The external biased magnetic field $B_0$ is along +$z$ direction. (b) The simulated (red symbol lines) and analytical (blue symbol lines) nonreciprocal transmission spectra of the YIG layer with $B_0$=0.05 T pointing in the +$z$ direction. $|t_{LL}^{+f}|^2$ and $|t_{RR}^{+f}|^2$ are the co-polarized transmission for the forward (-$z$-direction) normally incident LCP and RCP waves, respectively. (c) The transmission spectra of the metaisolator with $B_0$=0.05 T pointing in the +$z$ direction. $|t_{LL}^{+f}|^2$ and $|t_{RL}^{+f}|^2$ are the co- and cross-polarized transmission for the forward (-$z$-direction) normally incident LCP waves, and $|t_{RR}^{+b}|^2$ and $|t_{LR}^{+b}|^2$ are those for the backward (+$z$-direction) normally incident RCP waves. (d) The normalized multipole scattering power of the dielectric elliptical cylinder under the excitation of the forward-incident LCP wave. p, m, T, $Q_e$, and $Q_m$ are the electric dipole, magnetic dipole, toroidal dipole, electric quadrupole, and magnetic quadrupole, respectively. The inner image shows the corresponding electric and magnetic fields in the meta-atom at the resonant frequencies 5.5 GHz and 6.5 GHz.

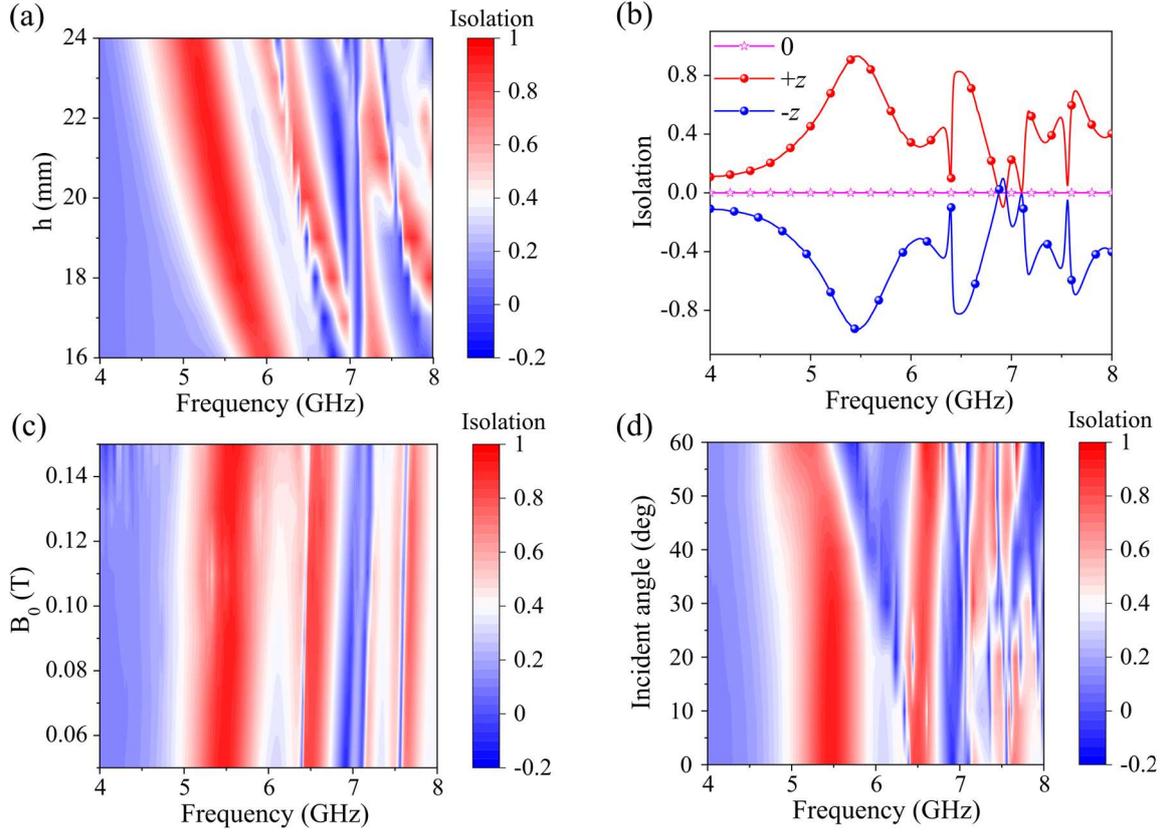

FIG. 2. The nonreciprocal characteristics of the PB-phase-based metaisolator. (a) The isolation ratio (i.e. $|t_{RL}^{+f}|^2 - |t_{LR}^{+b}|^2$) of the metaisolator as a function of the frequency and the height of the dielectric elliptical cylinder. (b) The isolation ratio of the metasurface with the ±z-biased magnetic field of 0.05 T or without the magnetic field for the normally incident CP light (i.e., $|t_{RL}^{\pm f}|^2 - |t_{LR}^{\pm b}|^2$). (c) The isolation ratio (i.e., $|t_{RL}^{+f}|^2 - |t_{LR}^{+b}|^2$) as a function of the external +z-biased magnetic field strength $B_0$ and the frequency of normally incident CP waves. (d) The isolation ratio (i.e., $|t_{RL}^{+f}|^2 - |t_{LR}^{+b}|^2$) as a function of the incident angle and the frequency of the CP waves. The external magnetic field is in the +z-direction with a magnitude of 0.05 T.

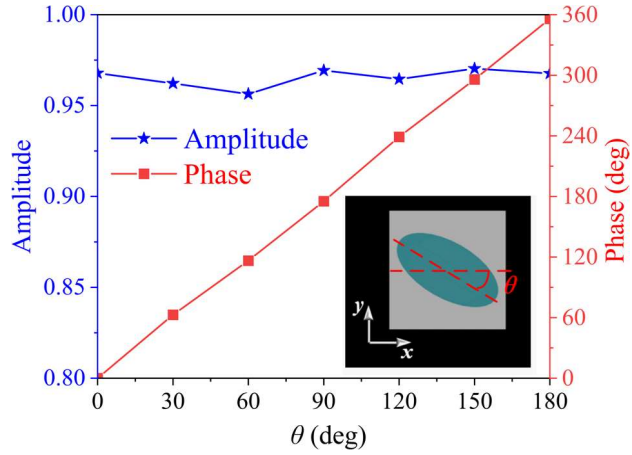

FIG.3. The cross-polarized transmission amplitude and phase shift for the metaisolator composed of elliptical dielectric cylinders with different orientation angles. The incident wave is LCP working at 5.5 GHz, and it normally incidents on the metasurface.

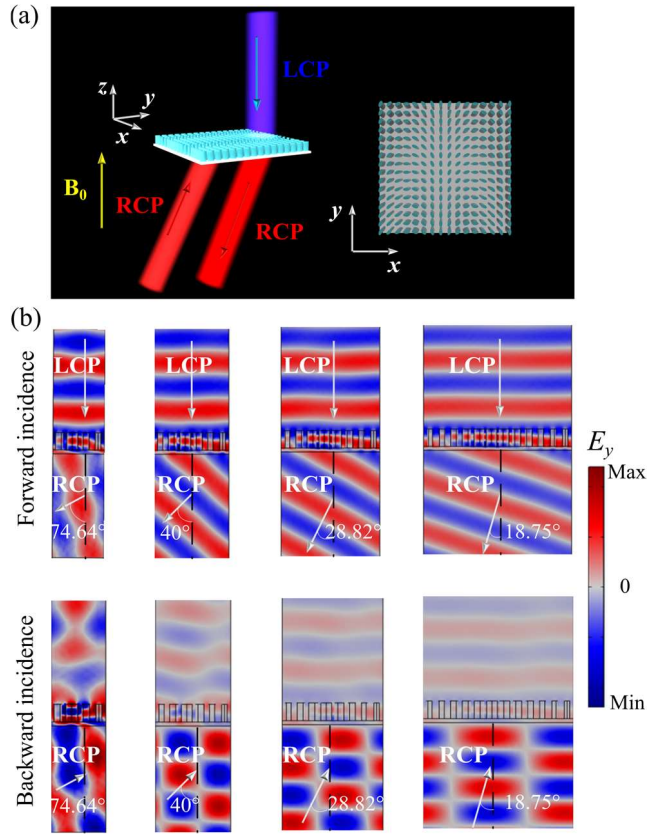

FIG. 4. Nonreciprocal metadeflector for beam steering. (a) The schematic for the nonreciprocal beam steering by the PB metadeflector with the magnetic-field biasing pointing in $+z$-direction. (b) The simulated normalized electric-field profiles for the supercells with different meta-atoms. The incident wave is LCP for the upper-row panels and RCP for the bottom-row panels, and their propagation directions are denoted by the white arrows. The frequency is at 5 GHz. The metadeflectors with the supercells consisting of 4, 6, 8, and 12 cells can achieve the deflection angle of 74.64°, 40°, 28.82°, and 18.75°.

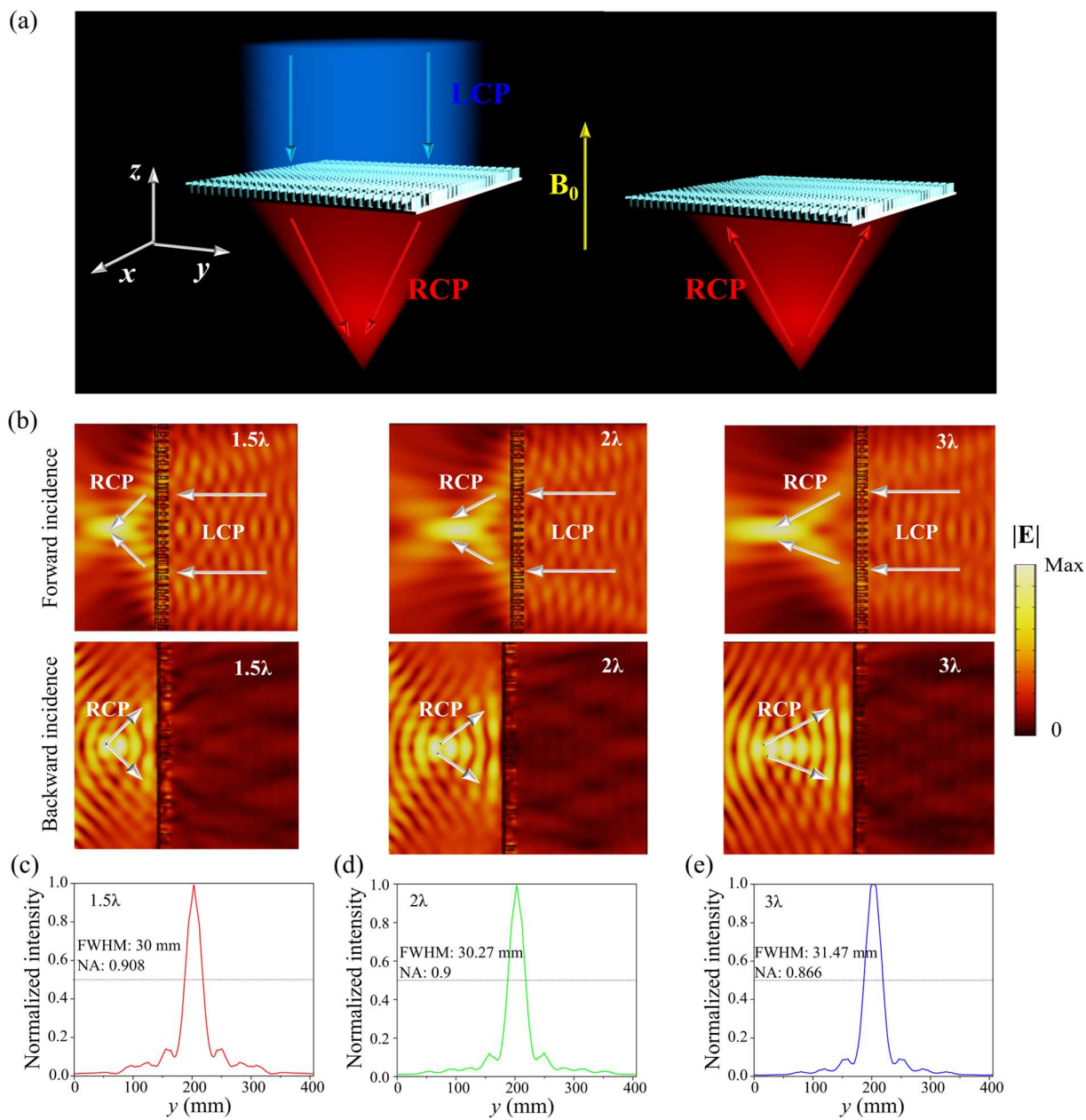

FIG. 5. The nonreciprocal PB metalens for beam focusing. (a) The schematic for the nonreciprocal focusing of the proposed PB metalens. The incident LCP beam is converted to the RCP beam and focused at one point, while the RCP radiation from the focal point cannot pass through the metalens. (b) The normalized electric-field distribution in the $yz$-plane when the forward incident LCP beam passes through the metalenses with different focal lengths $1.5\lambda$, $2\lambda$, and $3\lambda$ (the upper panels), and when the RCP wave radiated from the focal points backward propagates into the metalenses (the bottom panels). The normalized intensity distribution at the focal planes $z=-1.5\lambda$ (c), $z=-2\lambda$ (d), and $z=-3\lambda$ (e), respectively, corresponding to the three cases in the upper panels of (b).

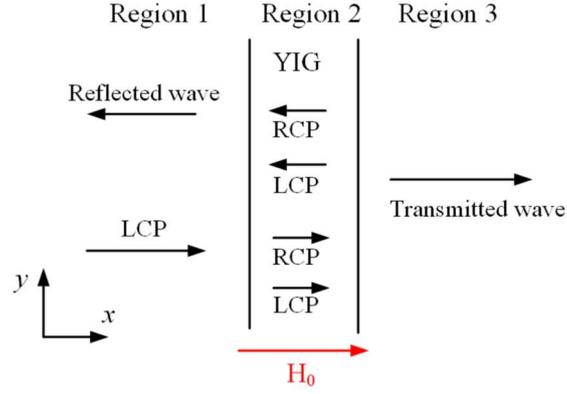

FIG. 6. The schematic for the thin YIG layer with +x-biased magnetic field under the normal incidence of a LCP plane wave.

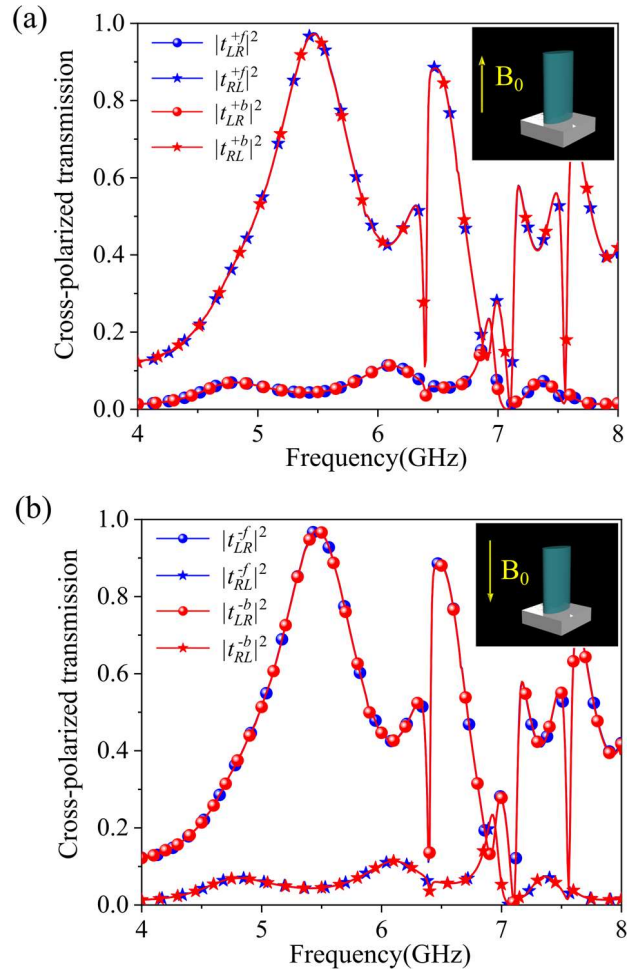

FIG. 7. The comparison of cross-polarized transmission of different CP waves with different propagation directions through the metaisolator with magnetic biasing along +z (a) and -z (b) directions shown in the insets. The magnitude of the magnetic field is 0.05 T. The subscript "$LR$" ("$RL$") represents the CP state conversion from RCP to LCP (LCP to RCP). The superscripts "+" and "−" indicate the +z- and −z-biased magnetic field; "$f$" and "$b$" stand for the forward (−z-direction) and backward (+z-direction) incidence.